\documentclass[pdflatex,sn-mathphys-num]{sn-jnl}

\usepackage{graphicx}%
\usepackage{multirow}%
\usepackage{amsmath,amssymb,amsfonts}%
\usepackage{amsthm}%
\usepackage{mathrsfs}%
\usepackage[title]{appendix}%
\usepackage{xcolor}%
\usepackage{textcomp}%
\usepackage{manyfoot}%
\usepackage{booktabs}%
\usepackage{algorithm}%
\usepackage{algorithmicx}%
\usepackage{algpseudocode}%
\usepackage{listings}%

\begin{document}

\title[Article Title]{Cortical Dynamics of Neural-Connectivity Fields}

\author*[1]{\fnm{Gerald K.} \sur{Cooray}}\email{gerald.cooray@ki.se}

\author[2]{\fnm{Vernon} \sur{Cooray}}\email{vernon.cooray@angstrom.uu.se}
\equalcont{These authors contributed equally to this work.}

\author[3]{\fnm{Karl J.} \sur{Friston}}\email{k.friston@ucl.ac.uk}
\equalcont{These authors contributed equally to this work.}

\affil*[1]{\orgdiv{Clinical Neuroscience}, \orgname{Karolinska Institutet}, \orgaddress{\street{Eugeniav}, \city{Stockholm}, \postcode{17177}, \country{Sweden}}}

\affil[2]{\orgdiv{Angstrom Laboratory}, \orgname{Uppsala University}, \orgaddress{\street{Lägerhyddsv 1}, \city{Uppsala}, \postcode{752 37}, \country{Sweden}}}

\affil[3]{\orgdiv{Functional Imaging Laboratory at Queens Square Institute of Neurology}, \orgname{University College London}, \orgaddress{\street{12 Queens Square}, \city{London}, \postcode{WC1N 3AR}, \country{United Kingdom}}}


\abstract{Macroscopic studies of cortical tissue reveal a prevalence of oscillatory activity, that reflect a fine tuning of neural interactions. This research extends neural field theories by incorporating generalized oscillatory dynamics into previous work on conservative or semi-conservative neural field dynamics. Prior studies have largely assumed isotropic connections among neural units; however, this study demonstrates that a broad range of anisotropic and fluctuating connections can still sustain oscillations. Using Lagrangian field methods, we examine different types of connectivity, their dynamics, and potential interactions with neural fields. From this theoretical foundation, we derive a framework that incorporates Hebbian and non-Hebbian learning — i.e., plasticity — into the study of neural fields via the concept of a connectivity field.}




\maketitle

\section{Introduction}

The cortical surface comprises layers of neural cells, scaffolded into a neural network by neural glia cells including astrocytes. Astrocytes also help maintain the homeostasis of neurotransmitters and extracellular electrolytes, and are considered to be involved in the transmission of membrane potentials \cite{araque2010glial}. Moreover, empirical studies of the cortex—using intracellular and extracellular electrodes—have revealed continuously fluctuating membrane potentials, leading to measurable extracellular potential gradients. Local field potentials, measured with macroscopic electrodes, exhibit oscillatory activity, while microelectrodes reveal spiking activity. The summation of extracellular potentials, resulting from neural firing (via axons) or slower dendritic postsynaptic changes, leads to oscillatory activity with frequency content ranging from approximately 1 Hz to 500 Hz \cite{einevoll2013modelling}. 

The biophysical pathway for potential propagation through the neuropil is varied and involves the transmission of action potentials along membranes, specifically via axons and dendrites, as well as between neurons through chemical and electrical synapses. The conductive properties of these pathways are highly dependent on the type of synapses involved. Chemical synapses, in particular, exhibit significant variability in the speed and direction of induced currents, which is influenced by transmitters, synaptic receptors, synapse-related metabotropic changes, and postsynaptic uptake dynamics \cite{gundelfinger2000molecular}. Additionally, current transmission across astrocytes is possible, though our understanding of this process remains limited \cite{araque2010glial}.

Theoretical models of the cortical surface and the induced dynamics have been developed by various authors, resulting in both intricate and simplified models. Intricate models incorporate many aspects of neural cells, allowing for a wide range of cellular dynamics, whereas simplified models include only a minimum set of neural components that capture the key characteristics of neuronal activity. Estimating model parameters for intricate models using data is unlikely to be effective due to the high dimensionality and complexity of the parameter space. However, these models can be useful for simulating data with specified network parameters and exploring the effects of parameter changes on dynamics. In contrast, simpler models have seen considerable development, particularly in the context of parameter estimation from data: a line of work initiated by Amari and others \cite{amari1974method,nunez1974brain,wilson1972excitatory,lopes1974model,wright1995simulation}.
For an in-depth review of neural field models, Cook et al provides a comprehensive analysis \cite{cook2022neural}.

The dynamics analysed in simpler models typically involve single or multilayer systems, with or without the constraint of oscillatory activity. Even within these simpler frameworks, it is possible to study the dynamics of neural cells interacting through action potentials. By carefully tuning these models, it has been shown that oscillatory activity can be achieved, often estimated as the average response of a neural population. While studying dynamics constrained to specific parameter settings might seem restrictive, it finds empirical support, as oscillatory activity is more prevalent than action potentials when recording cortical activity using macroscopic electrodes. In this study, we will explore the theoretical constraints required to produce oscillatory activity, for which there is still no complete understanding of the cellular or mesoscopic processes that generate them. However, self-organized criticality (SOC) provides a qualitative explanation for the prevalence of oscillatory dynamics. 

SOC induces a state of criticality in threshold-driven systems between order and disorder, as described and discussed in \cite{bak1988self,jensen1998self,friston2019free,jirsa2022entropy}. Experimental evidence suggests that neural dynamics are constrained by SOC, generating oscillations as observed in \cite{poil2012critical,shew2009neuronal}. Furthermore, several studies have suggested that systems near criticality have enhanced data compression capabilities; implying that natural forces drive systems toward optimizing data storage and manipulation \cite{ruffini2023structured}. Oscillatory activity, in this context, can be seen as a system exhibiting phase invariance, either globally or locally. In this paper, we expand on this idea, proposing that self-organized criticality, through its complex interactions, leads models to exhibit activity near points of invariance. The invariance structures of neural fields have been specifically analysed in the visual cortex and in non-biological neural networks \cite{bressloff2001geometric,poggio2016visual,bronstein2017geometric}. We demonstrate in the following sections how the cortical network induces invariance structures and how, under the assumption of local invariance, this supports a theory of neural connectivity interactions. While the spectrum of possible interactions is vast, the structural dynamics that allow oscillatory activity—or, in general terms, invariance structures—can be analysed using a concise set of scenarios. The global and local invariance of field models is well-studied in the dynamics of physical systems, including electromagnetism, gravity, and quantum field theory, where each theory is characterized by specific invariance structures.

In Section 2, we present the theoretical background on a model of the cortical surface, exploring the impact of connections on the dynamics and how these connections can generate a connection field. In Section 3, we review the biological substrate for the interaction between connectivity and neural activity and investigate some specific examples. In Section 4, we discuss the analysis presented in this paper, the biological context and empirical evidence.

\section{The cortical field}

In this section, we summarize the key findings presented in \cite{cooray2023NF}. The cortical surface is modelled as a thin sheet, just a few millimetres thick, covering an area of 1–2 $m^2$ in the human brain. Within this sheet, neural units span cortical layers, with connections either orthogonal to the surface (intrinsic connections) or running along the sheet (extrinsic connections), as depicted in Fig. 1. The cortical field dynamics emerge from multiple bi-layers of excitatory and inhibitory neural units, where balanced activation results in oscillatory activity. This balance allows the field to be expressed as a complex vector, with one complex component per bilayer. The dynamics are governed by equation 1. The first term on the right-hand side arises from the excitation-inhibition balance, leading to oscillations. The second term introduces perturbations to these oscillations when the interaction is weak. The matrix $\mathbf{W}$ represents the connectivity gain between different layers and between points $\mathbf{r}$ and $\mathbf{r}_0$ on the cortical surface. The vector field $\mathbf{S}$, in conjunction with $\mathbf{W}$, determines how field activity at point $\mathbf{r}$ influences the field activity at point $\mathbf{r}_0$. For a neural field, $\phi$, we have:

\begin{figure}[h]
\centering
\includegraphics[width=0.9\textwidth]{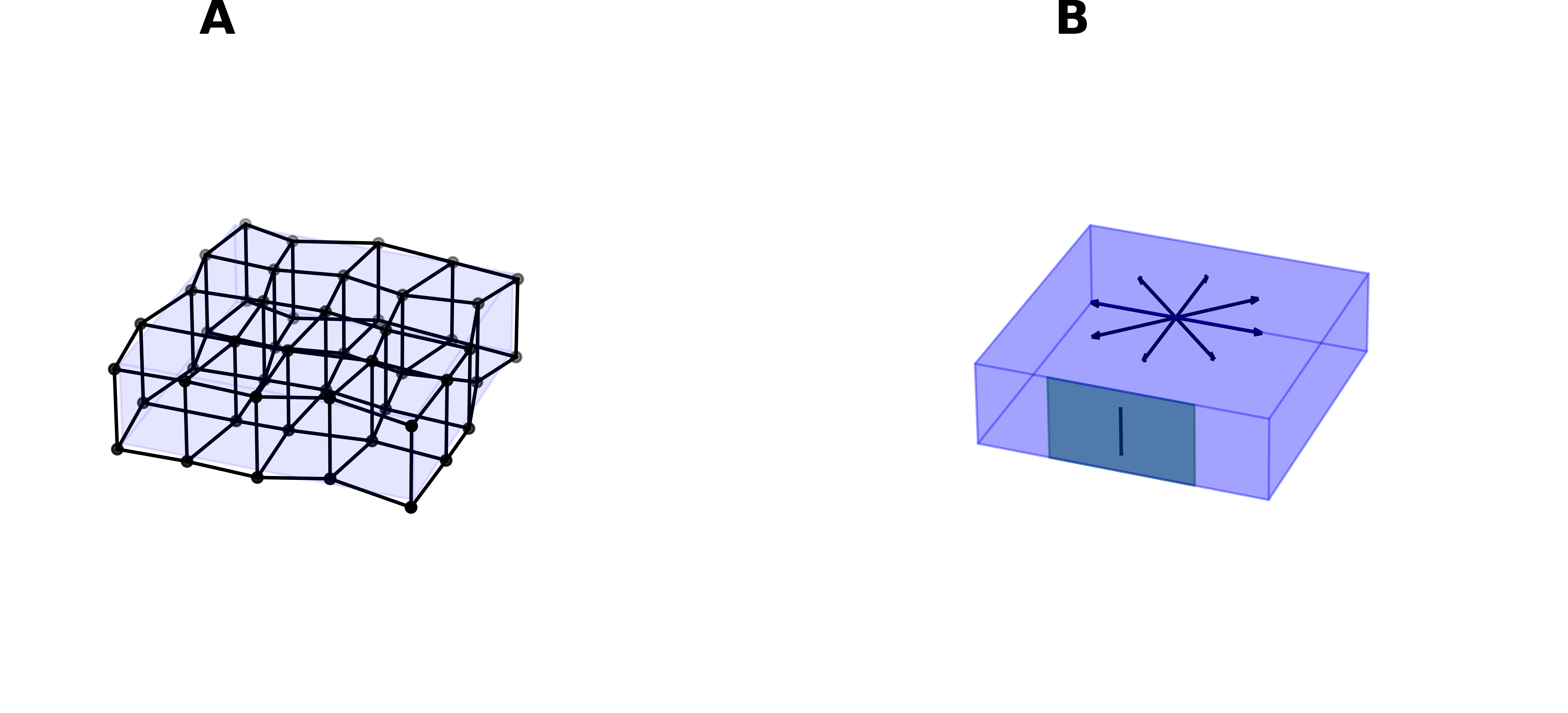}
\caption{A. The cortical sheet is modelled as a lattice of neural units, with excitatory and inhibitory units organized in two layers. Intrinsic connections link excitatory and inhibitory units between the two layers, while extrinsic connections link units within a layer. B. In the continuum limit, the model is depicted as a sheet. Extrinsic connections span different points on the sheet's surface, while intrinsic connections link varying depths within the cortical sheet (illustrated by an arrow in the shaded rectangle).}\label{fig1}
\end{figure}

\begin{equation}
\partial_{t} \boldsymbol{\phi} = -i\boldsymbol{\phi} + \iint_A \mathbf{W}(\mathbf{r}_0,\mathbf{r}) \, \mathbf{S}(\boldsymbol{\phi} (\mathbf{r}_0),\boldsymbol{\phi}(\mathbf{r})) \, \mathrm{d}A
\end{equation}

The integral covers a disc-shaped region around $\mathbf{r}_0$ on the cortical surface, where local connections to $\mathbf{r}_0$ are assumed to be relevant. The literature suggests that this region corresponds to the size of a few cortical columns, with a diameter ranging from 0.1 to 1 mm \cite{katzner2009local}. The neural field descriptions derived from the dynamical system in equation 1 have been extensively discussed in the literature \cite{cook2022neural,jirsa2009neural,coombes2005bumps}. We will assume that our system dynamics remain close to an equilibrium state and focus on perturbations around this state, implying that variations in neural field strength are small. Additionally, we will assume a long-wavelength approximation, excluding the high-frequency aspect of the spectral activity. In human brain recordings, frequencies below 100 Hz are usually considered to satisfy this approximation \cite{miller2009power}. In sections 2.2-5, we will explore the invariant structures of the neural field and their implications for dynamics. We will further generalize equation 1 to investigate connectivity between neural units and the related invariant structures, enabling us to propose an interaction between cortical connections and neural field activity.

\subsection{Dynamics for a Single Cortical Layer}
The dynamics of a single bi-layer are defined in equation 1, and under a long wavelength approximation, these dynamics can be expressed using a partial differential equation; see \cite{cooray2023NF} for further details.

\subsubsection{Dynamics of Weak Connectivity}
It is important to note that we are perturbing around a zero neural field state with weak connection strength, $W$. 
\begin{equation}
  \begin{aligned}
    |\phi| & \ll 1 \\
    & \\
    |\mathbf{W}| & \ll 1 \\
  \end{aligned}
\end{equation}
The \textit{S}-field is approximated by the first-order expansion of a sigmoid function, specifically $\tanh{\phi}$, see equation 1. To ensure that the connection parameter remains real, we define the connection as $-2iW = U$. While imaginary connection gains might seem unusual, they result from using a complex-valued dynamical equation and represent cross-connections between excitatory and inhibitory cells. This network configuration is responsible for generating oscillatory activity. By including only linear terms of $U$ and expanding the integral using a Taylor series for the integrand, we obtain a wave equation.
\begin{equation}
  \begin{aligned}
  \partial_{t}^{2} \phi & = -i\partial_{t}\phi + \partial_{t}\iint_A W\phi\ \, \mathrm{d}A\\
    \partial^{2}_{t} \phi & = -\phi +  \iint_A U \phi\, \mathrm{d}A \\
    \partial^{2}_{t} \phi & = -\phi +  \iint_A \left(U \phi\right)|_{r_0} \mathrm{d}A + \iint_A \frac{1}{2}\partial^{2}_{i}\left(U\phi\right)|_{r_0} (r-r_0)^{2} \, \mathrm{d}A\\
    & = -\phi + Ua\phi +\partial_{i}^{2}\left(U\phi\right)b\\
    & = - \left(1-Ua-b\partial_{i}^{2}U\right)\phi+Ub\partial_{i}^{2}\phi\\
    \end{aligned}
\end{equation}
 $a$ and $b$  are defined using integrals over the disc, $A$. 
\begin{equation}
  \begin{aligned}
    a & = \iint_A \, \mathrm{d}A\\
    b & = \frac{1}{2}\iint_A r^{2} \, \mathrm{d}A\\
  \end{aligned}
\end{equation}
The dynamical equation governing the system is given by equation 5. Note that equation 3 does not include first-order derivatives of $U$, reflecting the assumption that there is no intrinsic directionality of the connections on the cortical surface. This assumption can be relaxed, and there is some empirical evidence to do so; however, we will not include an analysis of such a structure as the derivations would be too cumbersome to follow. 
\begin{equation}
  \begin{aligned}
    \partial^{2}_{t} \phi - Ub\partial_{i}^{2}\phi& = - \left(1-Ua-b\partial_{i}^{2}U\right)\phi\\
    \end{aligned}
\end{equation}
Equation 4 is the Klein Gordon field (wave equation with mass), where the mass, $m$, and speed, $c$, terms are defined as follows which has been studied in \cite{cooray2023NF}. 
\begin{equation}
  \begin{aligned}
    m^{2}c^{4} & =  \left(1-Ua-b\partial_{i}^{2}U\right)\\
    c^{2} & = Ub\\
  \end{aligned}
\end{equation}
For the moment, we will assume that $U(r,r)$ is constant across the surface while approximating the system dynamics using $U$ and its higher-order derivatives (which we assume are even-dimensional). Additionally, we can derive the time derivatives of the dynamical system by integrating over the retarded time rather than the evaluation time simplifying the derivation as shown in equation 7. 

\begin{equation}
  \begin{aligned}
    \iint_{A} U(t_{ret},r) \phi(t_{ret},r) \, \mathrm{d}A  & = Ua\phi +  \iint_A  \frac{1}{2}\left[ -\frac{1}{Ub}\partial^{2}_{t}\left(U\phi\right)|_{r_0}+\partial^{2}_{i}\left(U\phi\right)|_{r_0}\right](r-r_0)^{2}\mathrm{d}A \\
    & = \left(Ua+b\partial_{i}^{2}U\right)\phi +  \iint_A  \frac{1}{2}\left[ -\frac{1}{b}\partial^{2}_{t}\left(\phi\right)|_{r_0}+U\partial^{2}_{i}\left(\phi\right)|_{r_0}\right](r-r_0)^{2}\mathrm{d}A \\
  \end{aligned}
\end{equation}
We can combine equation 3 and 7 to give an integral equation which will be equivalent to equation 3 where, $A'$ indicates that the integral is done over the retarded time. Equation 8 is a concise description of the neural field dynamics.  
\begin{equation}
  \begin{aligned}
\iint_{A'} U \phi \, \mathrm{d}A' &  = \phi \\
     \end{aligned}
\end{equation}

We now take a short detour—describing the Lagrangian formalism—as this will play an important part later in section 2. The neural field dynamics (equation 3 and 6) has an associated Lagrangian density which will be a mixture of quadratic functions of the neural fields and its derivatives, equation 9.
\begin{equation}
\mathcal{L}_0(\phi,\phi^{*},\partial_{t} \phi,\partial_{t} \phi^{*}) = \partial_{t} \phi^{*} \partial_{t} \phi - c^{2} \nabla \phi^{*} \nabla \phi + m^{2}c^{4} \lvert \phi \rvert^{2} 
\end{equation}
The notation can be simplified further—using standard terminology from classical field theory—by introducing a metric, $\mathbf{g}$. At present we are assuming the Minkowski metric which will not add any new information to the dynamical equation we have derived. We further use shorthand notation from physics, were repeated up and down indices (co-variant and contra-variant indices) are summed over; roman indices run from 1 to 2 and Greek indices from 0 to 2. The Lagrangian density is then given by equation 10.
\begin{equation}
\mathcal{L}_{0}(\phi,\phi^{*},\partial_{t} \phi,\partial_{t} \phi^{*}) = g^{\eta\mu}\partial_{\eta} \phi^{*} \partial_{\mu} \phi + m^{2}c^{4} \lvert \phi \rvert^{2}
\end{equation}
The metric defined by the connection $U$ is actually a pseudo-metric as it contains both positive and negative terms along the diagonal, equation 11. 

\begin{equation}
  \begin{aligned}
    {\mathbf{g}}^{\mu\nu} = \begin{pmatrix}1 & 0 & 0\\0&-Ub&0\\0&0&-Ub\end{pmatrix}
     \end{aligned}
\end{equation}
The dynamical equation of the neural fields (equation 6) is given by the variation of the neural field in the Lagrangian density. Varying the complex conjugate of the neural field, $\phi^*$, will give equation 6 and varying the neural field will give the complex conjugate of equation 6. This variation is given by the Euler-Lagrange equations. 

\subsubsection{Dynamics of Non-weak Connectivity}
The analysis in section 2.1.1 assumed that the interactions are weak; however, it is possible to drop this assumption. A system with strong intrinsic connections can be shown to have a non-zero stable state in contrast to the dynamics analysed in section 2.1.1 but the overall field equations can be shown to be similar. To formalise the above, we will start by defining our model using equation 12. The main difference between equation 2 and 12 is the strong intrinsic connection we have included in the form of a Dirac function $\delta$, i.e. the intrinsic connectivity is of much greater strength than the extrinsic and is modelled using a point like function.
\begin{equation}
  \begin{aligned}
  \partial_{t}\phi| & = -i\phi + \int T(\phi)W \\
  T(\phi) & = i\left(\alpha\phi+\beta\phi|\phi|^2\right)\delta (\mathbf{r}) +i \gamma\phi\\
  W & = |W|
   \end{aligned}
\end{equation}

The dynamical equation will be given by equation 13 and can be shown to have a set of non-zero stable states.
\begin{equation}
  \begin{aligned}
  \partial_{t}\phi & = -iW\left(1-\alpha + a\gamma + \beta |\phi|^2\right)\phi + iW\gamma b \nabla^2\phi\\
  \end{aligned}
\end{equation}
Perturbing equation 13 at any of these stable states will give (in a long wavelength approximation) a dynamical equation for a complex field (equation 14).
\begin{equation}
  \begin{aligned}
  \partial_{t}^2\left(\psi\right) & =  2W\beta\gamma b |\phi_0|^2\nabla^2  \left(\psi\right)\\ 
  \partial^2\psi & =  0\\ 
  c^2 & = 2W\beta\gamma b |\phi_0|^2\\
  \end{aligned}
\end{equation}
This state will have a different metric to what was defined in section 2.1.1 (as the speed of propagation of the neural fields is different) and moreover, there is an absence of a mass term.
\begin{equation}
  \begin{aligned}
  g^{\mu\nu}\partial_{\mu} \partial_{\nu}\psi & =  0\\ 
  \end{aligned}
\end{equation}
The metric, $\mathbf{g}$, is given in equation 16. 
\begin{equation}
  \begin{aligned}
    {\mathbf{g}}^{\mu\nu} = \begin{pmatrix}1 & 0 & 0\\0&-2W\beta\gamma b |\phi_0|^2&0\\0&0&-2W\beta\gamma b |\phi_0|^2\end{pmatrix}
     \end{aligned}
\end{equation}

The neural-connectivity field coupling as discussed in section 2.2-4 will be different for the neural field derived in 2.1.1 and that in section 2.1.2. Equation 5 is Lorentz invariant while equation 15 will be conformally invariant (which includes Lorentz invariance). The neural field described in section 2.1.1 will model small oscillation activity while that of 2.1.2 can be used to model activity showing high amplitude oscillations, e.g. alpha oscillations in human EEG \cite{nunez1974brain}.

\subsection{Single Layer Cortical Dynamics with Variable Intrinsic Connection}
In deriving the above equations we have assumed that the connection $U(r,r)$ in the cortical bi-layer is isotropic and does not change with time. However, there is compelling empirical evidence that cortical connections change along the surface and that they are time dependent, see section 3.

\begin{figure}[h]
\centering
\includegraphics[width=0.9\textwidth]{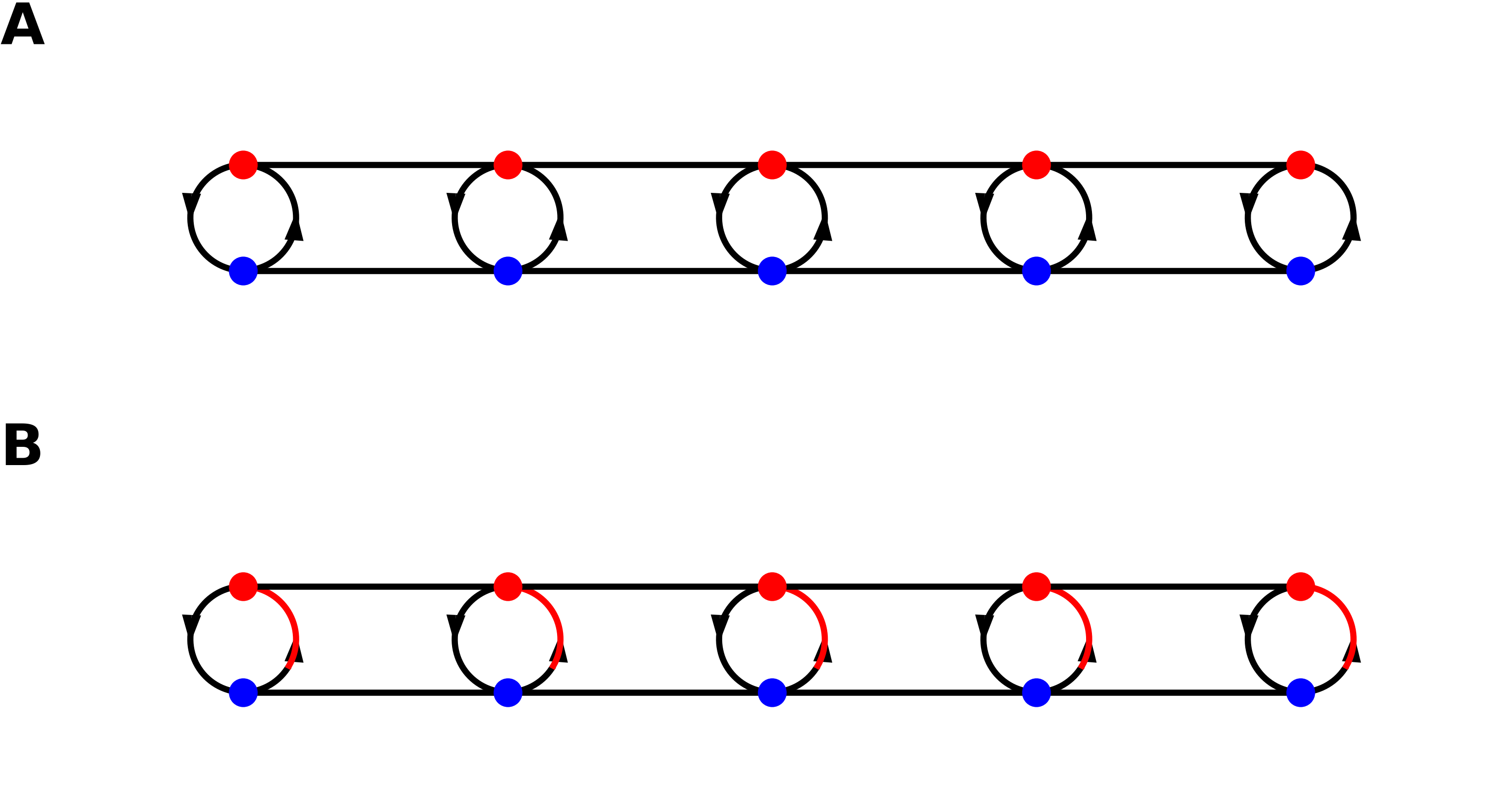}
\caption{Schematic figure of the bi-layer cortex with excitatory and inhibitory neural units. A. The bilayer has constant connections between the two layers, tuned to create oscillations, The connections are drawn as black circles between the layers. B. The bi-layer requires a compensatory variable connectivity to maintain oscillations. These variable connections are drawn as red arrows between the layers}\label{fig2}
\end{figure}

In this section we will analyse the effect of spatial and temporal variation of the intrinsic connection. See figure 2 for a schematic how variations in the intrinsic connection strength can be compensated by a variable connection strength to retain the oscillation as will be analysed in this section. In brief, the connectivity  $U$ above will be modulated using a space and time dependent phase, $\theta$. 
\begin{equation}
  \begin{aligned}
    U & \rightarrow  e^{i(\theta-\theta_{0})}U\\
  \end{aligned}
\end{equation}
The modulation of $U$ given by $\theta$ is a function on the cortical surface that varies with time and space. Including the modulated connection term into our dynamical system (eq. 8) gives us equation 18. 
\begin{equation}
  \begin{aligned}
        \phi & = \iint_{A'} e^{i(\theta^{'}-\theta)} U \phi^{'} \, \mathrm{d}A'\\
     \end{aligned}
\end{equation}
Expanding the integral and taking into account the retarded time gives the following equation.
\begin{equation}
  \begin{aligned}
\phi  &  =  \iint_{A'} Ue^{i(\theta^{'}-\theta)}\phi ^{'}\, \mathrm{d}A'\\
&  = aU\phi +b\partial_{i}^2U\phi  -e^{-i\theta}g^{\mu\nu}\partial_{\mu}\partial_{\nu} \left( e^{i\theta}\phi \right)\\
     \end{aligned}
\end{equation}
Equation 19 is a constraint on the time-space dependent connectivity field, $e^{i\theta}$, and the neural field, $\phi$. We will define a new variable, $\mathbf{A}$, and a perturbation constant, $\epsilon$, (to use standard nomenclature from classical field theory).
\begin{equation}
-\epsilon A_{\mu}=\partial_{\mu}\theta
\end{equation}
Combining equation 19-20—and after some manipulation—we get an expression for the interaction between the cortical and connectivity field in the format of a classical gauge field. We will define the connectivity field to be that of, $\mathbf{A}$, and the neural field to be equal to $\phi$.
\begin{equation}
  \begin{aligned}
-(1- aU\phi -b\partial_{i}^2U)\phi&  =   g^{\mu\nu}\left(\partial_{\mu}\partial_{\nu}\phi+\left(-i\epsilon\partial_{\mu} A_{\nu}-\epsilon^2 A_{\mu} A_{\nu}\right)\phi
 -2i\epsilon A_{\mu}\partial_{\nu}\phi\right)\\
 -m^{2}c^{4}\phi & = g^{\mu\nu}\left(\partial_{\mu}-i\epsilon A_{\mu}\right)\left(\partial_{\nu}-i\epsilon A_{\nu}\right)\phi\\
     \end{aligned}
\end{equation}
The terms in the parenthesis can be seen as derivative terms and are defined in classical field theory to be the \textit{covariant derivative}. 
\begin{equation}
\begin{aligned}
 D_{\eta}& = \partial_{\eta}-i\epsilon A_{\eta}\\
\end{aligned}
\end{equation}

Using the terminology defined we can succinctly rewrite the dynamics as shown in equation 23. 
\begin{equation}
\begin{aligned}
 g^{\eta\mu}D_{\eta}D_{\mu}\phi + m^{2}c^{4} \phi  & = 0\\
\end{aligned}
\end{equation}
Note that the modulation of the connection, $\theta$, does not give rise to solenoidal vector fields when the derivative is taken (as it is taken to be a real scalar).  However, we can allow for more variability in the connection giving us $A_{\eta}$-terms that are mixtures of gradient and solenoidal fields. Returning to equation 18 we define the modulation of the connection $U$ as follows, where $\mathbf{r}$ is at the centre of the integrating domain.
\begin{equation}
  \begin{aligned}
     \iint_A e^{i(\theta-\theta_{0})}U\phi\ \, \mathrm{d}A & \rightarrow i\iint_A e^{i(f(\mathbf{r}, z,z^*)-f(\mathbf{r}, 0,0))}U\phi\ \, \mathrm{d}A\\
  \end{aligned}
\end{equation}
When we derive the dynamic equations we get a similar expression but with the connectivity field defined using equation 25. 
\begin{equation}
  \begin{aligned}
       -\epsilon A_{\mu} & =\partial_{\mu}f(\mathbf{r}_0, z,z^*)\\
  \end{aligned}
\end{equation}
Note that this is not a gradient field as $f$ can be seen as a two dimensional vector. Equation 23 defines the interaction between the neural field and itself and the interaction between the neural and connectivity field; however, the interaction between the connectivity field and itself is not defined. The dynamics can be modified to include an interaction term for the connectivity field by including a kinetic interaction. This is done without too much complication by modifying the Lagrangian of the dynamical equation. The Lagrangian density for the unmodified system is given in equation 26 corresponding to the dynamical equation (equation 23). 
\begin{equation}
\mathcal{L}_{1}(\phi,\phi^{*},\mathbf{D} \phi,\mathbf{D} \phi^{*}) = g^{\eta\mu}D_{\eta} \phi^{*} D_{\mu} \phi + m^{2}\lvert \phi \rvert^{2} 
\end{equation}
To this Lagrangian density we will add a kinetic term for the connectivity field which will give us squared differential terms in the dynamical equations of motion. The Lagrangian density for the connectivity self-interaction is given in equation 27.

\begin{equation}
\mathcal{L}_{2}(A, \partial A) = \frac{1}{4} F_{\mu\nu}F^{\mu\nu}
\end{equation}
Where $F_{\mu\nu}$ is given in expression 28.
\begin{equation}
    \begin{aligned}
F_{\mu\nu}  & = \partial_{\mu}A_{\nu}-\partial_{\nu}A_{\mu}\\
    \end{aligned}
\end{equation}
The generalised Lagrangian density with interaction terms and self-interaction terms between neural fields and connectivity fields are given (equation 29) as the sum of the Lagrangian densities (eq. 26 and 27). 
\begin{equation}
\mathcal{L}(\phi,\phi^{*},\mathbf{D} \phi,\mathbf{D} \phi^{*}, A, \partial A) = g^{\eta\mu}D_{\eta} \phi^{*} D_{\mu} \phi + m^{2}\lvert \phi \rvert^{2} + \frac{1}{4} F_{\mu\nu}F^{\mu\nu} 
\end{equation}
Before we develop the theory any further, we will pause to discuss some important aspects of the above dynamical equation. 

1) The dynamics of the connectivity field will persist even when the neural field is 0. This partly surprising aspect of the theory finds support in experimental evidence where self-interacting connectivity fields have been described \cite{minerbi2009long,shimizu2021computational}. 

2) The Lagrangian density that was derived (equation 29) is the simplest one that allows for the dynamical equation to be invariant to phase transformations (as evidenced in data from cortical tissue) and with a dynamical self-interaction term for the connectivity field. 

3) The mode of interaction is causal on the speed of propagation of the neural and connectivity field between different events. The first term on the right-hand side of equation 29 mediates the causal dependency of the neural fields and the third term that of the connectivity field. If these fields were not included (e.g. if the Lagrangian density of equation 26 was used) we would have a situation where neural fields would create a disturbance in the connectivity field, which then would transmit infinitely quickly causing secondary changes in the neural field. The result would be neural fields propagating (indirectly) infinitely quickly, this again contrary to experimental findings. 

4) The equations are completely analogous to that of electromagnetism. Notice that there is no clear physical relation between classical electrodynamics and neural-connectivity field theory except the co-existence of phase invariance of oscillatory activity in both types of fields. However, in contrast to electromagnetism where the gauge field (the equivalent of the connectivity field) does not have an empirical counterpart (at least for classical fields), it does so for cortical dynamics. The empirical counterpart is given by the actual connections between the neural units via axons, dendrites, synapses and possibly through the astrocytic scaffolding in which the neural units are embedded. 

The main characteristics of cortical activity that allowed us to derive equation 29 rest on the following three criteria (i) cortical activity seems to fine tune the excitatory and inhibitory dynamics allowing for phase invariance (cf., excitation-inhibition balance). (ii) The existence of a neural field interacting with a connectivity field and (iii) a minimally complicated dynamical equation supporting point 1 and 2. Point (i) is supported by the oscillatory dynamics seen in data and also finds support in the idea of SOC, where the dynamics tend to a state of invariance, in the current setup a phase-invariance (SU(1)-invariance). Point (ii) does again have some empirical evidence. Point (iii) is mainly for computational reasons but is important, especially in situations where model parameters might be inferred from data.

\subsection{Cortical Surface Geometry and the Neural-Connectivity Field }
The geometry of the cortical surface—of relevance to the Neural-Connectivity field—is not determined by the physical geometry of the cortical tissue, but by the interaction of the fields on the surface on which they are defined. The geometry will be defined by the extrinsic connectivity. See figure 3 for a schematic depiction of this.
\begin{figure}[h]
\centering
\includegraphics[width=0.9\textwidth]{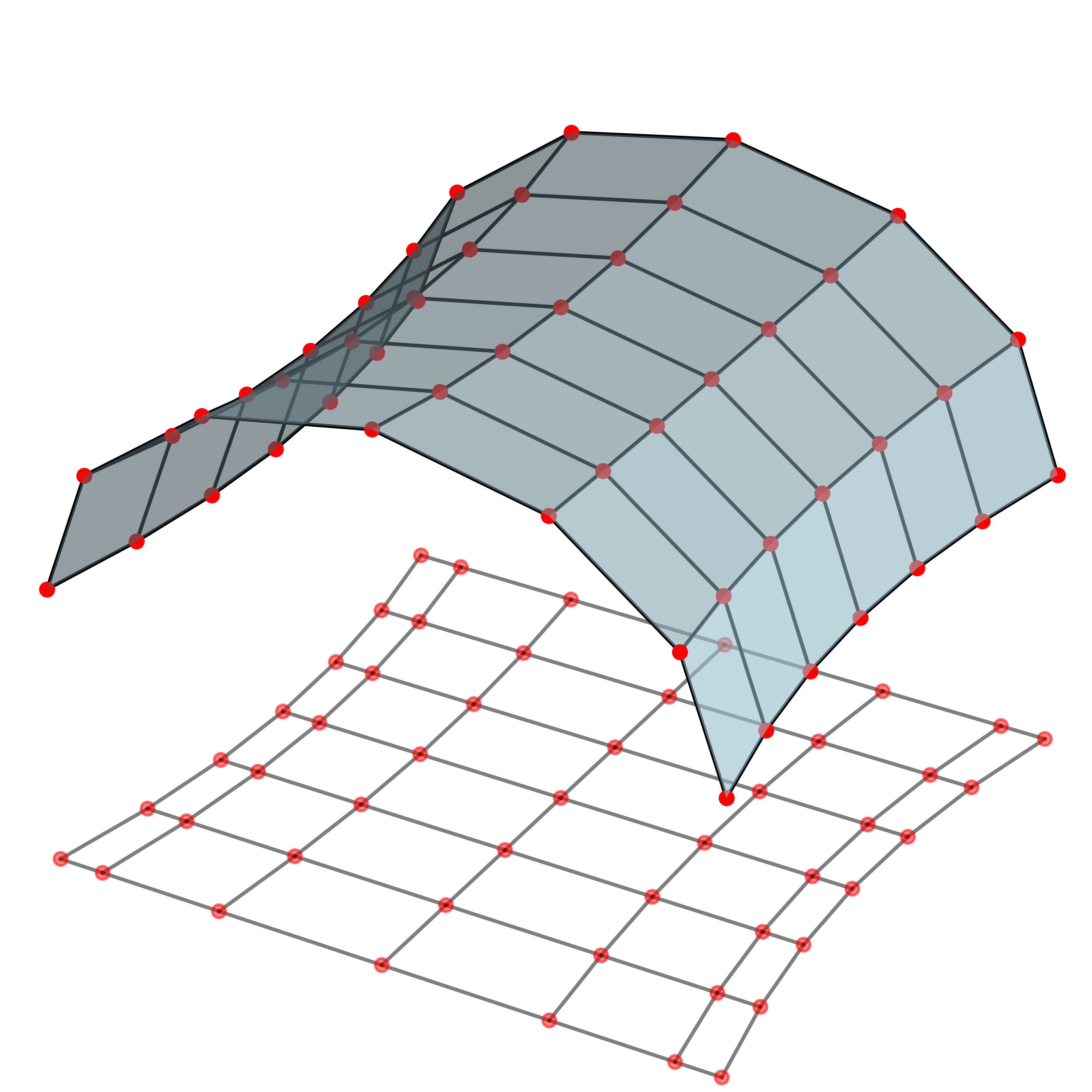}
\caption{Schematic figure of a lattice where the connection strength is reflected by the length of the connections. This lattice can be projected onto a surface such that each connection becomes of the same length as shown on the curved hyperbolic surface. This surface can be estimated from the connection strengths as discussed in this section}\label{fig1}
\end{figure}

In the following, we consider the mathematical constraints that fulfil a set of plausible geometrical assumptions about the neural connectivity model. 1) The cortical sheet is locally a smooth two-dimensional surface, i.e., a 2 dimensional differential manifold. 2) The concept of distance and angles exist which would require a Riemannian (or pseudo-Riemannian) manifold. 3) Movement of the neural fields on the surface is constrained by the intrinsic curvature of the surface without any external forces or fields required.

To model the interaction between the extrinsic connectivity and the geometry we will modulate the afferent connection (U) using the expansion in equation 30. 
\begin{equation}
  \begin{aligned}
\iint_{A}  U \phi \, \mathrm{d}A &  \rightarrow f( U,\mathbf{V},\mathbf{W} ,\phi) \\
     \end{aligned}
\end{equation}
The function $f( U,\mathbf{V},\mathbf{W},\phi)$ is defined in equation 31 where the connectivity consists of a scalar, a vector and a matrix component. These act on the neural field, the derivative and a second derivative of the neural field. 

\begin{equation}
    \begin{aligned}
        f(U,\phi) & = aU\phi - V^{\mu}\partial_{\mu}\phi - W^{\mu\nu}\partial_{\mu}\partial_{\nu}\phi\\
    \end{aligned}
\end{equation}
It can be shown that the geometry or equivalently the metric induced by the above second order expansion of the neural field and its derivatives is, in the simplest case, given by the following relations for the coefficients $\mathbf{W}$ and $\mathbf{V}$.
\begin{equation}
    \begin{aligned}
        W^{\mu\nu} & = g^{\mu\nu}\\
        V^{\nu} & = \frac{1}{\sqrt{|g|}}\partial_{\mu}\left(\sqrt{|g|}g^{\mu\nu}\right)\\
    \end{aligned}
\end{equation}
The function $f$ is given in equation 33 for the simplest geometrical setting.

\begin{equation}
    \begin{aligned}
        f(g^{\mu\nu},\phi) & = aU\phi - \frac{1}{\sqrt{|g|}}\partial_{\mu}\left(\sqrt{|g|}g^{\mu\nu}\partial_{\nu}\phi\right)\\
    \end{aligned}
\end{equation}
The dynamical equation is given by the following expression which simplifies as follows.

\begin{equation}
  \begin{aligned}
\phi = \iint_{A}  U \phi \, \mathrm{d}A  & = aU\phi - \frac{1}{\sqrt{|g|}}\partial_{\mu}\left(\sqrt{|g|}g^{\mu\nu}\partial_{\nu}\phi\right)  \\
 0 & = (1-aU)\phi + \frac{1}{\sqrt{|g|}}\partial_{\mu}\left(\sqrt{|g|}g^{\mu\nu}\partial_{\nu}\phi\right)  \\
     \end{aligned}
\end{equation}
The last equation is the Klein Gordon field on a geometry with metric defined by $g^{\mu\nu}$ and the Levi-Civita connection. The expansion for the connection function is constrained by the two dependent terms $V$ and $W$. This is necessary to define both the metric and the unique torsion free connection (the Levi Civita connection). With this we have fulfilled constraints 2 and 3 above. With this geometry the movement of waves can be defined by the metric itself and its derivatives (as can be seen in equation 34). The curvature of the surface will bend the waves according to equation 34. Without the constraint between $V$ and $W$ we would have defined both a metric and connection which would in general also create a torsion "force" on the surface. Constraint 3—i.e. the absence of external forces bending the fields as they propagate over the surface—can be relaxed but the dynamical theory will then need further elaboration. This is something we will not further analyse in this paper.

Equation 34 gives the interaction of the neural field with itself and between the connectivity field (i.e. the metric) and the neural field. Similar to the analysis in section 2.2. we can generalise the equation of motion to include the interaction of the connectivity field (the metric) with itself. This is easiest done by analysing the Lagrangian density of the dynamics. As it stands, the density is given in equation 35.
\begin{equation}
\mathcal{L}(\phi,\phi^{*},D_{\mu} \phi,D_{\mu}\phi^{*}) = \sqrt{g}\left(g^{\eta\mu}D_{\eta} \phi^{*} D_{\mu} \phi + m^{2}\lvert \phi \rvert^{2}\right)\\
\end{equation}
The simplest kinetic term that could be added to the Lagrangian is the Ricci scalar $R$ \cite{sotiriou2010f}. The value of the Ricci scalar at a point on the surface will measure the dispersion of straight lines originating from that point, or equivalently the curvature of the surface.  The Ricci scalar is defined using the metric, $\mathbf{g}$, as shown in equations 36.
\begin{equation}
    \begin{aligned}
        \Gamma^{\rho}_{\mu\nu} & = \frac{1}{2}g^{\rho\lambda}\left(g_{\lambda\mu , \nu}+g_{\lambda\nu , \mu}-g_{\mu\nu,\lambda}\right)\\ 
        R^\rho_{\sigma\mu\nu} & = \partial_{\mu}\Gamma^{\rho}_{\nu\sigma} -\partial_{\nu}\Gamma^{\rho}_{\mu\sigma}+\Gamma^{\rho}_{\mu\lambda}\Gamma^{\lambda}_{\nu\sigma}-\Gamma^{\rho}_{\nu\lambda}\Gamma^{\lambda}_{\mu\sigma}\\
        R & = g^{\mu\nu}R_{\mu\nu} = R^\lambda_{\mu\lambda\nu} \\ 
     \end{aligned}
\end{equation}
The modified Lagrangian with interaction terms between the connection terms (i.e. the metric) and the neural fields are given in equation 37. 
\begin{equation}
\mathcal{L}(\phi,\phi^{*},D_{\mu} \phi,D_{\mu}\phi^{*},R) = \sqrt{|g|}\left(g^{\eta\mu}D_{\eta} \phi^{*} D_{\mu} \phi + m^{2}\lvert \phi \rvert^{2} + R\right)\\
\end{equation}
This equation is also known as the Hilbert action (or the Einstein-Hilbert action) and will in 3 spatial dimensions (with 1 time dimension) give the field equations for general relativity. In 2 spatial dimensions the interaction between the neural and connectivity field will cause non-trivial dynamics. Using the dynamical equation derived from equation 36 we will show that the proposed theory inherently incorporates Hebbian and non-Linear Hebbian learning (section 3).

\subsection{Multi-layer Dynamics }
We now expand our study of the single (bi-) layer with several interacting layers. To obtain a theory which allows for analytical study we can perturb the connectivity around a state with only self-connections for each bi-layer and show that the perturbation of this (trivial) connectivity will give non-trivial dynamics. The connection between neural units gives the interaction between the different cortical layers and is a matrix varying over the cortical surface (n × n matrix for n-cortical layers) and for non-interacting layers it reduces to the identity matrix. We can modulate this matrix using a unitary matrix, $\mathbf{S}$, using the fact that the Lagrangian (eq. 10) is invariant to unitary transformations.  

\begin{equation}
  \begin{aligned}
    \iint_{A}U \phi& \rightarrow \iint_{A}\mathbf{S}^{-1}(r_0) \mathbf{S}(r)U\phi\\
  \end{aligned}
\end{equation}
Note that the perturbation is 0 at $r_0$ but varies with $r$. The dynamical equation will be given by a generalisation of equation 8, where we integrate over retarded time.
\begin{equation}
    \begin{aligned}
         \mathbf{\phi} & = \iint_{A'}U\mathbf{S}^{-1}(r_0) \mathbf{S}(r)\vec{\phi}\\
    \end{aligned}
\end{equation}
Expanding equation 39 and using the same steps as in section 2.2 we get the dynamical equation 40.
 \begin{equation}
    \begin{aligned}
       0 &=  g^{\mu\nu}\partial_{\mu}\partial_{\nu}\left(\mathbf{S}\vec{\phi}\right)-m^2\mathbf{S}\vec{\phi} \\
       0 & = \left( g^{\mu\nu}\partial_{\mu}\partial_{\nu}-m^2 \right)\left(\mathbf{S} \vec{\phi}\right)\\  
    \end{aligned}
\end{equation}
Writing out the expansion of $\mathbf{S}$ and the derivative terms shows that eq. 40 is not invariant to transformations but gives us instead relation 41. 
\begin{equation}
    \begin{aligned}
0 & = \mathbf{S}g^{\mu\nu}\partial_{\mu}\partial_{\nu}\vec{\phi} + 2g^{\mu\nu}\partial_{\mu}\mathbf{S}\partial_{\nu}\vec{\phi} - m^2\mathbf{S}\vec{\phi}\\
    \end{aligned}
\end{equation}
Similar to the analysis in section 2.2, we can introduce a connectivity field, or gauge field, to compensate for the 3 terms in equation 41. The covariant derivative given by equation 42 will compensate for the terms in relation 41. 
\begin{equation}
    \begin{aligned}
    D^{\mu} & = \partial_{\mu}-(\mathbf{S}^{-1}\partial_{\mu}\mathbf{S})\\
    \end{aligned}
\end{equation}
The dynamical field is invariant to transformations given by $\mathbf{S}$ as shown with the "commutation" relation in equation 43.
\begin{equation}
    \begin{aligned}
       \left( g^{\mu\nu}\partial_{\mu}\partial_{\nu}-m^2 \right)\left(\mathbf{S} \vec{\phi}\right) & =\textbf{S}\left( g^{\mu\nu}D_{\mu}D_{\nu}-m^2 \right)\vec{\phi} \\  
    \end{aligned}
\end{equation}
The dynamics of the neural and connectivity field interaction is given in equation 44.  
\begin{equation}
    \begin{aligned}
        g^{\mu\nu}\left(D_{\mu}D_{\nu}+m^2\right)\vec{\phi} & = 0\\
\end{aligned}
\end{equation}
The Lagrangian density is given in equation 45 and is invariant to unitary transformations. 

\begin{equation}
\mathcal{L}(\vec{\phi},\vec{\phi}^{*},\mathbf{D} \vec{\phi},\mathbf{D} \vec{\phi}^{*}) = g^{\mu\nu}D_{\mu} \vec{\phi}^{*} D_{\nu} \vec{\phi} + m^{2}\lvert \phi \rvert^{2}\\
\end{equation}

The differential operator, $\mathbf{D}$, can be parametrised for unitary transformations using a similar set-up that we used in section 2.2. For a N-layered cortical surface, we have N generators $\mathbf{T^a}$ for special unitary modulations (transformations) and the covariant derivative is defined in equation 46. 
\begin{equation}
    \begin{aligned}
D_{\mu} & = \partial_{\mu}-ig\mathbf{A}_{\mu}\\
    \end{aligned}
\end{equation}
Where $\mathbf{S}$ and $\mathbf{A}_{\mu}$are given by the following expressions (where we are summing over the repeated \textit{a} index).
\begin{equation}
    \begin{aligned}
\mathbf{S} & = e^{-i\mathbf{T}^a\theta^a}\\
g\mathbf{A}_{\mu}  & = \partial_{\mu}\theta^a\mathbf{T}^a\\
    \end{aligned}
\end{equation}
The connectivity field is defined to be that of $\mathbf{A}$. The above expression states explicitly that $\mathbf{A}_{\mu}$ is derived from a gradient field. However, like the analysis of single bi-layer dynamics, we can introduce solenoids fields in $\mathbf{A}_{\mu}$ by expanding the variation of the connectivity field. We will again parameterise the variation using a complex variable and its conjugate, instead of a real parameter as in equation 47 (i.e. $\theta \in \mathbb{C} \cong \mathbb{R}^2$).  The Lagrangian density does not allow for self interaction of the connectivity field and the simplest term that can be added  to introduce a self interaction of the connectivity field is the minimal coupling for the connectivity field. This minimal coupling  is given by the curvature of the connection field $\mathbf{A}_{\mu}$ (equation 48).

\begin{equation}
    \begin{aligned}
\mathbf{F}_{\mu\nu}^{a}  & = \partial_{\mu}A_{\nu}^{a}-\partial_{\nu}A_{\mu}^{a}+ g\sum{f^{abc}}A_{\mu}^{b}A_{\nu}^{c}\\
    \end{aligned}
\end{equation}
The term pre-multiplying the connectivity field, $f^{abc}$, is determined by the type of transformations that the connectivity field generates, in this case it is the structure constants of the Lie Algebra $SU(n)$.  The full Lagrangian density will be given by equation 49. 

\begin{equation}
    \begin{aligned}
         \mathcal{L} & = D^{\mu} \vec{\phi^*}D_{\mu} \vec{\phi} + m^{2}\vec{\phi^*}\vec{\phi} -\frac{1}{4}\mathbf{F}_{\mu\nu}^{a}\mathbf{F}^{a\mu\nu}\\
    \end{aligned}
\end{equation}
These are highly non-linear equations but are the least complex that ensure unitary invariance of the dynamics.

In summary, the above equations (46-49) constrain the interaction between the cortical layers. The validity of the constraints can only be established through experimental cortical recordings in animals and humans. The analysis presented in section 2.2-2.3 can at least partly be validated by measuring cortical activity using macroelectrodes, e.g. subdural grid recordings in animals and humans; as this will allow a relatively dense sampling of the cortical surface within the size of the grid (about 20-25 $cm^2$).  However, section 2.4 analyses the constraints on the intrinsic connectivity of the proposed theory. A dense sampling of the intra-laminar dynamics is needed to validate this theory. There has been growing interest in intralaminar dynamics after the advent of several new electrode types allowing for ultra-dense sampling of the cortical column, e.g., neuropixel probes \cite{jun2017fully}.  In section 3, we will discuss the neurobiological substrate for a variable connectivity field and its interaction with the neural field. We also investigate specific dynamical systems described in section 2 and their predictions on connectivity dynamics.

\section{Connectivity Dynamics}
In section 2 we gave a detailed analysis of possible interaction fields for neural and connectivity fields that followed the criteria we set-up: 1) A self-tuned dynamical system with oscillations or a generalisation of oscillations, 2) Interaction between the neural and connectivity field and 3) the "simplest" dynamics fulfilling criteria 1) and 2). In this section we present the neurophysiological framing of an interacting neural-connectivity field in the cortex fulfilling criteria 1 and 2, we also discuss some specific examples in section 3.3. 

\subsection{Criteria 0 - The Multilayered Cortical Sheet}
The cortex has been assumed to be a multilayered sheet. There is a histological basis for this, as layers are seen in microscopic analysis of the cortical surface \cite{brodmann1909vergleichende}. The neural and connectivity field; however, are functional aspects of the cortical surface which might not map to its structural organisation. Laminar studies of the cortical sheet have showed evoked responses with a gradual change in LFP signals as the cortex is transversely sampled using multielectrodes, where 2-3 distinct patterns are noted throughout the cortical sheet. The electrophysiological evidence would suggest that the cortex could be divided into 2 or 3 functional layers \cite{chien2023laminar,fishman2000complex,schroeder1998spatiotemporal,haegens2015laminar}. We have in section 2 given the dynamics for single layer (one bi-layer) and multi layer dynamics. The corresponding symmetries for 1-3 functional layers are the unitary groups U(1), SU(2) and SU(3). The symmetries result in oscillations for a single bi-layer, U(1), and what we will define as a generalisation of oscillations for SU(2) and SU(3). We will analyse specifically "generalised" oscillations in a neural-connectivity field with SU(2) invariance in section 3.3. 

\subsection{Criteria 1 - Oscillations and Generalised Oscillations}
The first reported study of oscillatory activity from the cortex was the 8-10 Hz oscillations noted by Hans Berger using scalp electrodes placed on the skull \cite{berger1929eeg}. The existence of oscillations has been shown to be ubiquitous in cortical media and found in many species and in different locations of the cortex as described in \cite{buzsaki2004neuronal}. The frequency content ranges from below 1 Hz up to about 500 Hz. The power content is typically decreasing with increasing frequency with a sharp increase in gradient around 100 Hz which has been modelled using a neural field similar to that described in section 2.1 in \cite{cooray2023NF,miller2009power}. Even though oscillations are highly prevalent in electrophysiological recordings of cortical tissue, measuring them is a complex procedure and care must be taken not to misinterpret the presence or absence of them in empirical data.

A detailed description of methods used to characterise oscillatory behaviour in cortical recordings is given in \cite{donoghue2022methodological}. It is well known that that the generation of oscillations in cortical tissue is context specific.  An intricate interaction between interneurons, pyramidal cells, chemical and electrical synapses has been shown to generate oscillations in the hippocampus \cite{whittington2003interneuron,winfree1980geometry, glass2001synchronization}. Similar interactions would probably generate oscillations seen in other cortical structures, e.g., gamma oscillations in the prefrontal cortex \cite{jung2021neuronal}. The complexity of the cortical substrate for oscillations has been further elucidated in a range of studies foregrounding the importance of dendrite dynamics in the generation of LFPs \cite{sinha2022active}. The origin of the neural oscillations is multifaceted as described above; however, they all induce a phase invariant symmetry (U(1)) in the dynamics, the mathematical substrate for oscillations. The intense research that has been done in mammalian visual systems starting with the work of \cite{hubel1968receptive} has revealed cortical activity with flavours, or hints of higher invariant structures in the visual system, including SU(2) invariance \cite{bressloff2001geometric,ermentrout1979mathematical,baroni2023learning,cadena2019deep}. 

\subsection{Criteria 2 - Neural-Connectivity Interaction and Hebbian Dynamics}
We have constructed neural-connectivity field theories for different complexities of the cortical connections, section 2. This interaction between neural activity and connectivity is similar to Hebbian learning which is often considered to rest on N-methyl-D-aspartate channel (NMDA-R) dynamics \cite{artola1987long}. The mathematical framework for Hebbian-like learning has been suggested to be constrained by relations similar to that of equation 50 \cite{gerstner2002mathematical}.
\begin{equation}
  \begin{aligned}
  \frac{dw}{dt} = Cv_a v_b
  \end{aligned}
\end{equation}
The connection gain between the two connected neural units a and b is given by w and the activity by $v_a$ and $v_b$.  In the following subsections we will show how Hebbian and non-Hebbian learning is inherent in the dynamics of the neural-connectivity fields.

\subsubsection{Single Layer Intrinsic Dynamics and Hebbian Learning}
In section 2.2 we derived the neural-connectivity field of the single layered cortex governed by equation 23.  There will be a conserved current $J_{\mu}$ associated with the Lagrangian density defined in equations 51. 
\begin{equation}
  \begin{aligned}
  J_{\mu} & =-i\left(\phi^{*}\partial_{\mu}\phi-\partial_{\mu}\phi^{*}\phi\right)-2\epsilon A_{\mu}\phi\phi^{*} \\
  \partial^{\mu}J_{\mu} & = 0 \\
 \end{aligned}
\end{equation}
We will assume that the coupling between the neural and connectivity field, $\epsilon$, is weak allowing us to further simplify equation 51.
\begin{equation}
  \begin{aligned}
  J_{t} & = -i\left(\phi^{*}\partial_{t}\phi-\partial_{t}\phi^{*}\phi\right) = -|\phi|^{2}\\
  \end{aligned}
\end{equation}
Using the self-interaction for the connectivity field we get an expression for the dynamic interaction between the connectivity field and the conserved current.
\begin{equation}
  \begin{aligned}
  J_{\nu} & = \partial^{\mu}F_{\mu\nu} = \partial^{\mu}\left(\partial_{\mu}A_{\nu}-\partial_{\nu}A_{\mu}\right)\\
  J_{t} & = \partial^{\mu}\left(\partial_{\mu}A_{t}-\partial_{t}A_{\mu}\right)\\
  J_{t} & = \partial^{i}\partial_{i}A_{t}\\
 \end{aligned}
\end{equation}
The connectivity field can then be calculated as a function of the neural field.
\begin{equation}
  \begin{aligned}
   -|\phi|^{2} & = \partial_{i}^2 A_{t}\\
  \end{aligned}
\end{equation}
This expression can be recognised as Gauss law with the following solution for the connectivity field. 
\begin{equation}
  \begin{aligned}
   \partial_{t}f  = A_t & =-log(r)|\phi|^{2}\\
  \end{aligned}
\end{equation}
This expression indicates, for fixed $r$, Hebbian learning (compare with equation 50). In equation 51 we could include higher order terms in the expansion of the sigmoid function giving us the following expression for the connectivity field.
\begin{equation}
  \begin{aligned}
  J_{t} & = -|\phi|^{2} - a_2 |\phi|^{4} -  a_3 |\phi|^{6}- ...- \\
  \end{aligned}
\end{equation}
The expression for the rate of change of connection gain will then follow non-linear Hebbian dynamics.
\begin{equation}
  \begin{aligned}
   \partial_{t}f  = A_t & =-log(r)\left(|\phi|^{2} + a_2 |\phi|^{4} +  a_3 |\phi|^{6}+ ... +\right)\\
  \end{aligned}
\end{equation}

\subsubsection{Single Layer Extrinsic Dynamics and Hebbian Learning}
In the above derivation we investigated the interaction between the neural field and intrinsic connection as was studied in section 2.2.; however, the interaction between the neural field and the extrinsic connection brings more complexity. We now investigate a neural-connectivity model with a scalar field that varies only with time. The connectivity field dynamics are given in equation 58, where we have an expression linking the curvature structure of the cortical sheet (on the LHS) and the stress energy tensor of the neural field (RHS) which is derived from equation 37. 
\begin{equation}
    \begin{aligned}
        G_{\mu\nu} = R_{\mu\nu}-\frac{1}{2}R   & = kT_{\mu\nu} \\ 
    \end{aligned}
\end{equation}
The stress energy tensor is given in equation 59. 
\begin{equation}
    \begin{aligned}
        T_{\mu\nu}  = g_{\mu\rho}\frac{\partial \mathcal{L}}{\partial(\partial_{\rho}\phi)}\partial_{\nu}\phi -g_{\mu\rho}\delta^{\rho}_{\nu}\mathcal{L}\\ 
    \end{aligned}
\end{equation}
We will study the interaction between the $T_{\mu\nu}$ and the connectivity field given by the metric. We can use a similar procedure to that used when deriving the Friedman equations in general relativity as the symmetries of the two problems are similar \cite{thorne2000gravitation}. The Robertson Walker metric in 2 dimensions is given in equation 60, where \textit{a(t)} is an expansion component.
\begin{equation}
    \begin{aligned}
    ds^2 = dt^2 -a(t)^2 \left(\frac{dr^2}{1-kr^2} + r^2d\theta^2\right)
    \end{aligned}
\end{equation}
The curvature in a polar coordinate system is given in equations 61.  
\begin{equation}
    \begin{aligned}
        G_{tt} &  = \frac{k}{a^2} + \frac{a'^2}{a^2}  \\
        G_{rr} & = -\frac{aa''}{1-kr^2}\\
        G_{\theta\theta} & = -r^2aa''\\
     \end{aligned}
\end{equation}
We will, furthermore, simplify the stress energy tensor assuming that the scalar field does not have a specific direction. 
\begin{equation}
    \begin{aligned}
       T_{tt} & = - \sqrt{|g|}g_{tt}\left(g^{ii}\partial_{i} \phi^{*} \partial_{i} \phi - m^{2}\lvert \phi \rvert^{2}\right) = g_{tt}A\\
         T_{ii} & = - \sqrt{|g|}g_{ii}\left(g^{tt}\partial_{t} \phi^{*} \partial_{t} \phi +g^{jj}\partial_{j} \phi^{*} \partial_{j} \phi- m^{2}\lvert \phi \rvert^{2}\right) = -g_{ii}B\\
        \end{aligned}
\end{equation}
The first of equations 62, $T_{tt}$, will be given by a density field and the second a sum of a density and pressure field $T_{ii}$. The notation and terminology we use is the same as that used in the derivation of the Friedmann Equations. There will also be a conservation of the neural field current resulting in equation 63. 
\begin{equation}
    \begin{aligned}
        0 &  = \frac{d}{dt}\left(a^2\rho\right) + p\frac{d}{dt}\left(a^2\right)\\
    \end{aligned}
\end{equation}

The field equations (eq. 60) will then fix the a-parameter (equation 64).
\begin{equation}
    \begin{aligned}
        \frac{k}{a^2} + \frac{a'^2}{a^2} & = A\\
        \frac{a''}{a}  &= -B\\
       \end{aligned}
\end{equation}
Assuming a neural field state with high mass term we can simplify the above equation as shown in equations 65. 
\begin{equation}
    \begin{aligned}
        A = \frac{\rho_0}{a^2} & = \frac{k}{a^2} + \frac{a'^2}{a^2}\\
   a'^2 &  =  \rho_0  -k\\
   \frac{da}{dt} &  =\pm \sqrt{\rho_0  -k}\\
    \frac{da}{dt} &  =\pm \sqrt{m^2|\varphi|^2  -k}\\
    \frac{da}{dt} &  \approx \pm m|\varphi|\\
      \end{aligned}
\end{equation}
The last equation defines the relations between the change in the metric parameterised by \textit{a} and the neural field strength. This relation indicates non-linear Hebbian dynamics. 

\subsubsection{Multi-Layer Intrinsic Dynamics and Hebbian Learning}
We can now consider the multilayered cortex using similar steps to our analysis of the single layered cortex although the dynamics get more intricate due to the increased complexity of the invariance structures. The covariant derivatives are defined in equations 66, where $T^a$ is one of three generators of the invariance structure (SU(2))
\begin{equation}
  \begin{aligned}
   D_{\mu}\varphi & = \partial_{\mu}\varphi -ig T^{a} A^{a}_{\mu}\varphi\\
   \left(D_{\mu}\varphi \right)^* & = \partial_{\mu}\varphi^* +igA^{a}_{\mu}\varphi^* T^{a}\\
  \end{aligned}
\end{equation}
The current associated with the Lagrangian density (equation 49) is given in equation 67 and has increased complexity  in comparison to the single layer cortex as the generators ($T$) are included in the equations. Note that they do not in general commute with each other or the covariant derivatives.
\begin{equation}
  \begin{aligned}
   j_{\mu}^{a} & = -i\left[\varphi^*T^{a}D_{\mu}\varphi-(D_{\mu}\varphi)^*T^{a}\varphi \right]\\
  \end{aligned}
\end{equation}
The commutation relation between the generators is given in equations 68 together with their two dimensional representation (the Pauli Spin matrices). 
\begin{equation}
  \begin{aligned}
    \left[T^{a},T^{b}\right] &  = i\epsilon_{abc}T^{c}\\
  {\mathbf{T}}^{1} & = \begin{pmatrix}0 & 1 \\1&0\end{pmatrix}\\
  {\mathbf{T}}^{2} & = \begin{pmatrix}0 & -i \\i&0\end{pmatrix}\\
  {\mathbf{T}}^{3} & = \begin{pmatrix}1 & 0 \\0&-1\end{pmatrix}\\
 \end{aligned}
\end{equation}
The first generator, $T^1$, will map cross connection between excitatory neural units in layer 1(2) with inhibitory in layer 2(1). The second generator, $T^2$, will map cross correlations between excitatory neural units in layer 1(2) and excitatory neural units in layer 2(1). The last generator will map self-connection within each layer. The conserved currents associated with the dynamics are given in equation 69.
\begin{equation}
  \begin{aligned}
  j_{\mu}^{a} & = -i\left[\varphi^*T^{a}\partial_{\mu}\varphi-\partial_{\mu}\varphi^*T^{a}\varphi-ig\varphi^* T^{a}T^{b} A^{b}_{\mu}\varphi  -igA^{b}_{\mu}\varphi^* T^{b}T^{a}\varphi \right]\\
  j_{\mu}^{a} & = -i\left[\varphi^*T^{a}\partial_{\mu}\varphi-\partial_{\mu}\varphi^*T^{a}\varphi-ig\varphi^* T^{a}T^{a} A^{a}_{\mu}\varphi  -igA^{a}_{\mu}\varphi^* T^{a}T^{a}\varphi \right]\\
  j_{\mu}^{a} & = -i\left[\varphi^*T^{a}\partial_{\mu}\varphi-\partial_{\mu}\varphi^*T^{a}\varphi-2ig\varphi^* \varphi A^{a}_{\mu}\right]\\
  \end{aligned}
\end{equation}
The 3 currents do not interact; however, the generators create a rotation as will be seen below.  As the currents are conserved, we estimate the connectivity field as a function of the neural fields and the generators. 
\begin{equation}
  \begin{aligned}
  0 & = \left[\varphi^*T^{a}\partial_{t}\varphi-\partial_{t}\varphi^*T^{a}\varphi-2ig\varphi^* \varphi A^{a}_{t}\right]\\
   A^{a}_{t} & = \frac{\left[\varphi^*T^{a}\partial_{t}\varphi-\partial_{t}\varphi^*T^{a}\varphi\right]}{2ig\varphi^* \varphi }\\
   & = \frac{-\varphi^*T^{a}\varphi}{g\varphi^* \varphi }\\
  \end{aligned}
\end{equation}
The connectivity field is defined using the connection gain, $f_a$ as shown in equation 71.
\begin{equation}
  \begin{aligned}
  S & = e^{-if_a (r)T^a}\\
  A_{\mu}^a & = \partial_{\mu}f_a(r)\\
  \end{aligned}
\end{equation}
The change in connection gain is given as a function of the neural fields.
\begin{equation}
  \begin{aligned}
  \partial_{t}f_1(r) & = \frac{\varphi_{1}^*\varphi_{2}+\varphi_{2}^*\varphi_{1}}{g(\varphi_{1}^2 + \varphi_{2}^2)}\\
     \partial_{t}f_2(r) & = \frac{-i\varphi_{1}^*\varphi_{2} + i\varphi_{2}^*\varphi_{1}}{g(\varphi_{1}^2 + \varphi_{2}^2)}\\
     \partial_{t}f_2(r) & = \frac{\varphi_{1}^*\varphi_{1} - \varphi_{2}^*\varphi_{2}}{g(\varphi_{1}^2 + \varphi_{2}^2)}\\
  \end{aligned}
\end{equation}
All changes are seen to be real-valued and indicate non-linear Hebbian dynamics. The dynamical equation is given by equation 73. 
\begin{equation}
    \begin{aligned}
        0 & = D_{t}^{2}\vec{\phi}+m^2 \vec{\phi}\\
        0 & = g^{\nu\mu}\left(\partial_{\nu}+igA^a_{\nu}T^a\right)\left(\partial_{\mu}-igA^a_{\mu}T^a\right)\vec{\phi}+m^2 \vec{\phi}\\
        0 & = \left(\partial^{\mu}\partial_{\mu}-ig\partial^{\mu}A^a_{\mu}T^a+g^2 A^{a\mu}A^{b}_{\mu}T^{a}T^{b}\right) \vec{\phi}+m^2 \vec{\phi}\\
 \end{aligned}
\end{equation}
Using the fact that we are looking at field with only a temporal variation further simplification is possible giving us equation 74.
\begin{equation}
    \begin{aligned}
        0 & = \left(\partial^{2}_t-ig\partial_{t}A^a_{t}T^a+g^2 A^{at}A^{b}_{t}T^{a}T^{b}\right) \vec{\phi}+m^2 \vec{\phi}\\
        0 & = \left(\partial^{2}_t-ig\partial_{t}A^a_{t}T^a+3g^2 (A^{a}_{t})^2I\right) \vec{\phi}+m^2 \vec{\phi}\\
        0 & = \left(\partial^{2}_t-ig\partial_{t}A^a_{t}T^a\right) \vec{\phi}+\left(3g^2 (A^{a}_{t})^2+m^2\right)I\vec{\phi}\\
         0 & \approx \left(\partial^{2}_t-ig\partial_{t}A^a_{t}T^a\right) \vec{\phi}+m^2\vec{\phi}\\
 \end{aligned}
\end{equation}
The above equation is non-linear; however, to get an overall view of the dynamics we will assume that the second term in the parenthesis is constant. We can then solve the equation with the solution given in equation 75.
\begin{equation}
    \begin{aligned}
         0 & =  \partial^{2}_t \vec{\phi} + m^2 \left(I-i\frac{g\partial_{t}A^a_{t}T^a}{m^2}\right) \vec{\phi}\\
        \vec{\phi} & = e^{\pm im\left(I- i\frac{g\partial_{t} A^a_{t}T^a}{m^2}\right)^{\frac{1}{2}}t}\\
         \vec{\phi} & = e^{\pm \left(imI - \frac{g\partial_{t} A^a_{t}T^a}{2m}\right)t}\\
         \vec{\phi} & = e^{\pm imt}e^{\left(\mp i\frac{g\partial_{t} A^a_{t}T^a}{2m}\right)t}\\
         \vec{\phi} & = e^{\pm imt}e^{\mp i\alpha^a T^at}\\
\end{aligned}
\end{equation}

Where $\alpha$ is defined in equation 76. 
\begin{equation}
    \begin{aligned}
        \alpha & = \frac{g\partial_{t} A^a_{t}T^a}{2m}\\
\end{aligned}
\end{equation}

The last exponential in equation 75 is expanded and simplified in equation 77. 
\begin{equation}
    \begin{aligned}
        e^{- i\alpha T^a t} & = \sum\frac{1}{n!}\left(- i\alpha T^a t\right)^n\\
          & = \sum\frac{1}{2n!}\left(-1\right)^n \left(\alpha t\right) ^{2n} I+ \sum\frac{1}{2n+1!}(-i)(-1)^n\left(\alpha t\right)^{2n+1}T^a\\
          & =cos(\alpha t)\begin{pmatrix}
         1& 0\\0&1\end{pmatrix} +sin(\alpha t) \left(-iT^a\right)\\
         \end{aligned}
\end{equation}

The three generators will generate dynamics with "generalised" oscillations involving the two interconnected bilayers (equation 78).
\begin{equation}
    \begin{aligned}
        e^{- i\alpha T^1 t} & = \begin{pmatrix}
         cos(\alpha t)& -e^{i\frac{\pi}{2}}sin(\alpha t)\\e^{-i\frac{\pi}{2}}sin(\alpha t)&cos(\alpha t)\end{pmatrix}\\
         e^{- i\alpha T^2 t} & = \begin{pmatrix}
         cos(\alpha t)& -sin(\alpha t)\\sin(\alpha t)&cos(\alpha t)\end{pmatrix}\\
         e^{- i\alpha T^3 t} &  = \begin{pmatrix}
         e^{-i\alpha t}& 0 \\0&e^{i\alpha t}\end{pmatrix}\\
         \end{aligned}
\end{equation}
The generalised oscillations give modulations of the oscillating signals in the cortical layers. The first generator will give an oscillation of the amplitude between the layers together with modification of the phase. The second generator will give an amplitude oscillation between the layers and the third a phase modification within each layer. Activity of type 2) is noted in recordings of visual LFPs where the oscillating gamma signal is seen to move between layers after stimulation of the cortex. However, the neural connectivity field will give a combination of the above oscillations in the general case. 

\subsubsection{LFP Simulation in Multi-Layer Cortex}
The multilayered evoked LFP for a single stimulus can be estimated using equation 75. We will assume a 2 layered cortex with the neural-connectivity field with local SU(2) invariance. The stimulation will be given by a Dirac pulse on layer 1. 
\begin{equation}
    \begin{aligned}
        S(t) = \delta(t)\\
         \end{aligned}
\end{equation}
We will calculate the solution for a single layered cortical sheet (where the off-phase component due to the connectivity field is given by $C$ which is assumed much smaller than the mass term). 
\begin{equation}
    \begin{aligned}
        \left(\partial^2_t  + m^2  + iC\right)\phi = \delta(t)\\
        \end{aligned}
\end{equation}
Solving this equation using the Laplace transform will give expression 81.
\begin{equation}
    \begin{aligned}
       \int \left(\partial^2_t  +m^2 + iC\right)g(s)e^{-s t} & = \int e^{-s t}\\
       \int \left(s^2  +m^2 + iC\right)g(s)e^{-s t} & = \int e^{-s t}\\
        g(s) & = \frac{1}{s^2+m^2 + iC}\\
        \hat{g} & = H(t)\frac{1}{m}e^{imt}e^{-\frac{Ct}{2}}\\
        \end{aligned}
\end{equation}
Combining this with the generalised oscillation of the second generator (eq. 78) in the two layered cortex will give us a solution of a decaying wave moving between the layers, equation 82, see figure 2. The rotational speed between the layers, $\alpha$, is given in equation 76.  
\begin{equation}
    \begin{aligned}
         \vec{\phi(t)} & =  H(t)\frac{1}{m}e^{imt}e^{-\frac{Ct}{2}}\begin{pmatrix}
         cos(\alpha t)& -sin(\alpha t)\\sin(\alpha t)&cos(\alpha t)\end{pmatrix}\\
         \end{aligned}
\end{equation}
\begin{figure}[h]
\centering
\includegraphics[width=0.9\textwidth]{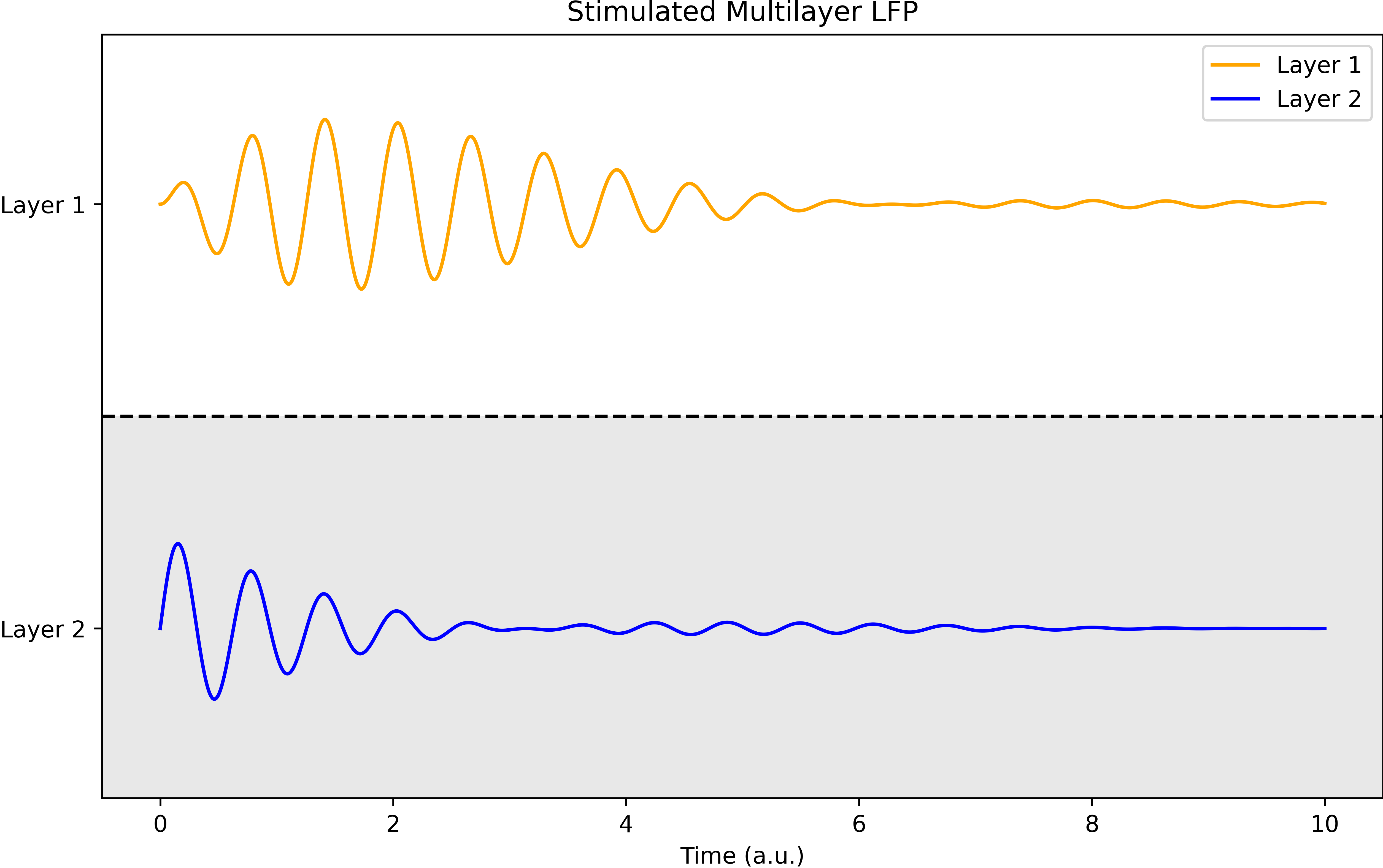}
\caption{This is a schematic figure illustrating the effect of generalised oscillations in a 2 layer cortical sheet. The response to a Dirac pulse in layer 2 is estimated where the induced wave moves from layer 2 to layer 1 before decaying.}\label{fig4}
\end{figure}
\section{Discussion}

In this study, we further develop the field model for a cortical surface described in \cite{cooray2023NF}. The field dynamics are primarily governed by a wave equation, specifically the Klein Gordon equation in its simplest form. All system interactions are wave-based, which imposes causality. As demonstrated in section 2.1, the concept of a retarded potential can be introduced to naturally encapsulate this causality, a theme that was introduced in section 2 and used throughout the paper. The derived wave equations possess several symmetries that constrain cortical connectivity, and breaking these symmetries would dampen the waves. The ubiquity of oscillations in cortical tissue suggests that many wave-preserving symmetries remain intact. In a dynamical system, symmetries can be global or local. Global symmetries seem less likely, as they would imply uniform connectivity constraints across the entire cortical surface. Instead, local constraints, which vary in time and position, appear more plausible. This idea can be reformulated as a gauge field theory, as shown in section 2.2, where a gauge field (or dynamic connectivity field) compensates for apparent asymmetries in the neural field. Importantly, we derived not a symmetric neural field but a symmetric neural-connectivity field. The neural connectivity field estimation was based on three principles assumed for cortical dynamics: 1) the system self-tunes to exhibit oscillations or generalized oscillations, 2) there is interaction between the neural and connectivity fields, and 3) the neural connectivity field is the simplest model satisfying the first two principles. The first two principles can be empirically tested, while the third is a computational assumption.

As outlined in section 3.2, oscillatory activity is prevalent in electrophysiological recordings of the cortex, observed across various cortical regions, such as the hippocampus and frontal neocortex. Neural oscillations are deemed so essential that the brain is thought to tune itself to generate them \cite{haken1992synergetics,buzsaki2006rhythms}. This concept is supported by self organized criticality, where dynamic systems reach a critical state, such as oscillation generation \cite{bak1988self,poil2012critical,shew2009neuronal}. While generalized oscillations, involving interactions between cortical layers, are challenging to demonstrate in tissue, multiple studies have detected high coherence between oscillatory signals across the laminar structure, along with movement of current dipoles, similar to the simulation we conducted in section 3.3.4 \cite{michelson2018isoflurane,swadlow2002activation,mitzdorf1987properties}. Additionally, studies have reported intricate invariances in neural field activity within the visual cortex, further supporting the notion of dynamics governed by invariant structures \cite{bressloff2001geometric}.

The dynamical interaction between neural activity and connection gain between different neural populations was first suggested by \cite{hebb2005organization} and later rephrased as \textit{neurons that fire together wire together} by \cite{lowel1992selection}. The interaction between the neural and connectivity fields is a core feature of Hebbian learning, a well-established mechanism in both mammalian and non-mammalian brains. This was supported by the discovery of long term potentiation (LTP) and depression (LTD) which has been seen in different regions of the mammalian brain including hippocampus, amygdala and neocortex but also in invertebrates \cite{bliss1973long,artola1987long,chapman1990long,stanton1989associative,dudek1993bidirectional,glanzman1995cellular}. Hebbian dynamics have been partly replaced by the mechanism of spike time dependent plasticity (STDP) \cite{markram1997regulation}. The sensitivity to the inter-spike duration for STDP has been estimated to lie between 5-50ms where changes within this range show opposite effects on the gain of the synapse depending on temporal order \cite{markram1997regulation,bell1997synaptic,debanne1998long,bi1998synaptic,magee1997synaptically}. The cellular basis for these dynamics, particularly LTP, LTD, and STDP, is thought to involve postsynaptic changes driven by slow NMDA-R dynamics. However, interactions between back-propagating action potentials, presynaptic NMDA-R, and voltage-gated calcium channels (VGCCs) have been shown to significantly shorten the timescale to about 5ms \cite{caporale2008spike}. Additional processes, such as synaptic augmentation and direct electrolytic changes, also have timeframes in the range of 1-10 ms, enabling rapid connectivity changes \cite{morgan2019kv1,hempel2000multiple}. Moreover, pyramidal cells exhibit linear changes in gain under the influence of parvalbumin (PV) cells in the visual cortex, with these rapid changes reflecting shifts in the connectivity field \cite{atallah2012parvalbumin}. There are several biological substrates that may contribute to fast-changing connectivity fields, with the specific mechanisms likely varying by cortical area but could all constitute to the connectivity field investigated in section 2 and 3.

\section{Conclusion}

We have analysed and formalized the interaction between a neural field and local symmetries, introducing the concept of a connectivity field. The complete theory describes an interacting neural-connectivity field, incorporating Hebbian dynamics as a core feature. Specific cases are explored, with predictions and simulations compared against empirical data, providing support for the general framework of a neural-connectivity field. However, further research is needed to validate these predictions on cortical activity with electrophysiological data obtained from intra-laminar recordings and dense cortical sampling.

\backmatter



\section*{Declarations}
\subsection*{Funding}
KF is supported by funding for the Wellcome Centre for Human Neuroimaging (Ref: 205103/Z/16/Z), a Canada-UK Artificial Intelligence Initiative (Ref: ES/T01279X/1) and the European Union’s Horizon 2020 Framework Programme for Research and Innovation under the Specific Grant Agreement No. 945539 (Human Brain Project SGA3). VC is supported by B. John F. and S. Andersson’s donation at Uppsala University.
\subsection*{Conflict of Interest/Competing Interests}
The authors have no competing interests to declare that are relevant to the content of this article.
\subsection*{Author Contribution}
Conceptualisation: G.K.C. Writing original draft: G.K.C. Writing: reviewing and editing: G.K.C., V.C. and K.F. All authors reviewed the manuscript.
Writing, reviewing and editing of submitted manuscript: GC, VC and KF.

\bibliography{sn-bibliography}

\end{document}